\documentclass[apj]{emulateapj}
\usepackage{amsmath}
\usepackage{amssymb}
\usepackage{graphicx}
\usepackage{epsfig}
\usepackage{apjfonts}
\usepackage{rotating}

\newcommand{\etal}{{ et~al.~}}

\newcommand{\ergscm}{\,{\rm ergs\,s$^{-1}$\,cm$^{-2}$}\,}
 
\newcommand{\gax }{{\lower0.8ex\hbox{$\buildrel >\over\sim$}}}
\newcommand{\lax }{{\lower0.8ex\hbox{$\buildrel <\over\sim$}}}

\shorttitle{Galactic S Stars}
\received{2011 April 29}
\begin{document}


\title{Galactic S Stars: Investigations of Color, Motion, and Spectral Features}
\author{Elizabeth Otto\altaffilmark{1}}
\email{otto.65@buckeyemail.osu.edu}
\author{Paul J. Green\altaffilmark{2}}
\author{and Richard O. Gray\altaffilmark{3}}

\altaffiltext{1}{The Ohio State University, Columbus, Ohio 43210 USA}
\altaffiltext{2}{Harvard-Smithsonian Center for Astrophysics, 60 Garden
 Street, Cambridge MA 02138}
\altaffiltext{3}{Department of Physics and Astronomy, Appalachian State
  University, Boone, NC 28608, USA} 

\begin{abstract}
Known bright S stars, recognized as such by their enhanced s-process
abundances and C/O ratio, are typically members of the asymptotic
giant branch (AGB) or the red giant branch (RGB).  Few modern digital
spectra for these objects have been published, from which intermediate
resolution spectral indices and classifications could be derived.  
For published S stars we find accurate positions using the Two-Micron
All Sky Survey (2MASS), and use the FAST spectrograph of the Tillinghast
reflector on Mt. Hopkins to obtain the spectra of 57 objects.  We make
available a digital S star spectral atlas consisting of 14 spectra of
S stars with diverse spectral features.  We define and derive basic
spectral indices that can help distinguish S stars from late-type (M)
giants and carbon stars.   We convolve all our spectra with the Sloan
Digital Sky Survey (SDSS) bandpasses, and employ the resulting $gri$
magnitudes together with 2MASS $JHK_s$ mags to investigate S star colors.
These objects have colors similar to carbon and M stars, and are
therefore difficult to distinguish by color alone.  Using near and
mid-infrared colors from IRAS and AKARI, we identify some of the stars
as intrinsic (AGB) or extrinsic (with abundances enhanced by past
mass-transfer).  We also use $V$ band and 2MASS magnitudes to calculate
a temperature index for stars in the sample.    We analyze the
proper motions and parallaxes of our sample stars to determine upper
and lower limit absolute magnitudes and distances, and confirm that
most are probably giants.  
\end{abstract}



\section{INTRODUCTION}
\label{sec:introduction}

When a low to intermediate mass star exhausts its core supply of
hydrogen, contraction and heating of the core causes the outer layers
of the star to expand and cool, whereby the star becomes a red giant
(RG) with an average effective temperature of 5000\,K or lower.  Once
the core temperature rises to about 3$\times 10^8$\,K, helium burning
begins, and the stellar surface contracts and heats creating the
horizontal branch (for Population II) or red clump (for Population I)
stars on the color-magnitude diagram.  Completion of core helium
burning and the start of helium shell burning returns the star to a
luminous, cooler phase as it evolves onto the asymptotic giant branch
(early- or E-AGB) where the star greatly expands and the hydrogen
shell is (almost) extinguished.  Later, as the He-burning shell
approaches the H-He discontinuity, a phase of double shell burning
begins.  Recurrent thermal instabilities in this thermally-pulsing
asymptotic giant branch (TP-AGB) phase are driven by helium shell
flashes, periodic thermonuclear runaway events in the He-shell
\citep{1983ARA&A..21..271I} The energy released by these pulses expand
the star, whereby the hydrogen shell is again basically extinguished
during some time. A third dredge up might then occur, and afterward
the hydrogen shell will re-ignite (and a new thermal pulse will
occur).  During these short-lived pulses, nucleosynthesis products
from combined H-shell and He-shell burning are dredged up to the outer
layers of the star.  The complex details of the dredge-up and its
outcome are strongly mass-dependent (\citealt{2005ARA&A..43..435H} and
references therein), but generally speaking, surface abundance
enhancements result in C, He, and the $s$-process elements, including
Ba, La, Zr, and Y.

S and Carbon (C) stars are traditionally thought to be members 
of the TP-AGB.  In addition to possibly elevated C/O ratios, these
stars exhibit other unique spectral characteristics as a result of the
dredge-up.  While M giants often exhibit titanium oxide
(TiO) absorption bands, both carbon and zirconium have a higher
affinity to free oxygen.  Therefore, a higher C/O ratio also results
in the disappearance of TiO bands and the appearance of zirconium
oxide (ZrO) bands.  S stars are therefore distinguished primarily by
their ZrO bands, while C stars show strong C$_2$ and  CN bands.
The S star spectral class is divided into three subtypes (in order of
increasing carbon abundance):  MS, S, and SC stars.  Classification
into one particular class is difficult, but is generally based on the
C/O ratio and the relative strengths of TiO, ZrO, and CN molecular
absorption bands.  Overall, a star in the S star class is generally
defined as having a C/O ratio (r) of $0.5 < r < 1.0$
and strong ZrO absorption features
\citep{1993A&A...271..463J,2010arXiv1011.2092V}.  Pure S stars are
those that show 
only ZrO bands, and no TiO bands \citep{1978MNRAS.184..127W}.
Classification is complicated by the fact that increased intensity of
the ZrO bands may be due to an excess of Zr, rather than being
exclusively tied to the C/O ratio \citep{2003IAUS..210P..A2P}. 
Furthermore, many AGB S and C giants are highly variable, and since 
the strength of molecular bands depends on the effective temperature
and surface gravity of the star, spectral type can vary significantly
with time. 

S stars whose spectra show lines of the short-lived element
$s$-process technetium (Tc) are known as 
"intrinsic" S stars, and are most likely be members of the TP-AGB.
$^{99}$Tc is only produced by the $s$-process, while the star is on the
AGB.  Since the half-life of $^{99}$Tc is only 2.13 x 10$^5$ years,
and the average duration of the AGB phase of stellar evolution is
roughly 1 Myr, stars exhibiting significant abundances of technetium
are almost certainly members of the AGB \citep{2000A&A...360..196V}.
By contrast, technetium-poor S stars are known as "extrinsic" S stars.
These stars are likely to be part of binary systems and their unusual
chemical abundances probably originated in a past mass-transfer
episode \citep{2000A&A...360..196V}.  The present extrinsic S star
likely once accreted $s$-process rich material from its companion, a
TP-AGB star at the time that has since evolved into a white dwarf.
Extrinsic S stars therefore display enriched $s$-process elements,
despite having never produced these elements themselves.  Most known
extrinsic S stars are assumed to be members of the red giant branch
(RGB).  They can sometimes be distinguished from intrinsic S stars on the basis
of color, because they do not necessarily show the red excess
characteristic of stars on the AGB.  While all currently known
extrinsic S stars are thought to be giants, these definitions open up
the possibility of extrinsic S stars on the main sequence (hereafter
referred to as dwarf S stars) that have previously accreted $s$-process
material and carbon from a companion but whose Tc has since
decayed. Analogous to the dwarf C stars \citep{Green91}, a faint S
star could be shown to be a dwarf if it has a sufficiently large
parallax or proper motion; a large proper motion predicts a
tangential velocity greater than the Galactic escape speed unless the
object is faint and nearby.   

S stars are very similar to C stars, with the major difference
being a slightly lower C/O ratio.  Faint high galactic latitude carbon
(FHLC) stars were once assumed be giant stars at large distances.  In
1977, however, G77-61 was discovered to have high proper motion and an
upper limit absolute magnitude of +9.6, therefore making it the first
known carbon dwarf star \citep{1977ApJ...216..757D}.  In this case,
the term dwarf refers to a star on the main sequence.  \citet{Green91}
discovered four other dwarf carbon (dC) stars using techniques 
that selected known carbon stars for high proper motion.  Now, well
over 100 dwarf carbon stars are known \citep{2004AJ....127.2838D}.
The local space density of dCs is far higher than all types of C
giants combined \citep{1992ApJ...400..659G}.  Therefore, contrary to
previous assumptions, dwarf C stars are the numerically dominant type
of carbon star in the Galaxy. 

Since dC stars are now known to be so common, we have begun a search
for dwarf S stars.  There is little reason to believe that dwarf S
stars should not exist.  In fact, they may be more abundant than dC stars, since
less carbon enhancement is required to produce an S dwarf as compared
to a C dwarf.  On the other hand, if the range of abundance ratios
that produces the characteristic features of S stars is 
narrow, they may be quite rare.  Furthermore, the range of C/O ratios
that produces S star spectral features may be different for giants and
dwarfs, given the higher gravities and larger temperature ranges in
dwarfs.   

We use medium-resolution digital spectra of known S stars (which are
probably giants) to generate corresponding Sloan Digital Sky Survey
(SDSS) colors, to see whether their colors might be sufficiently
distinctive to search for additional S stars in the SDSS with
reasonable efficiency.  We also use the FAST spectra to investigate
molecular band spectral indices for classification.  

Our paper is organized as follows:  
in \S\ref{sec:obs} we briefly review the observations and our data
reduction and processing procedures.  In section \S\ref{sec:spatlas}
we describe the production of a spectral atlas of S stars,
and derive spectral indices from the molecular band strengths in
\S\ref{sec:specind}. In
\S\ref{sec:colors} we describe the generation of SDSS colors and
analyze the overall colors of the FAST sample in order to determine
whether an efficient color selection for S stars exist.  In
\S\ref{sec:intext} we discuss techniques for distinguishing intrinsic
and extrinsic S stars and apply them to the FAST sample.  In
\S\ref{sec:motion} we analyze the parallaxes and proper motions of S
stars in the sample to characterize upper and lower limit magnitudes
and distances.  We use colors to generate approximate temperature
indices for S stars in \S\ref{sec:temp}.  In \S\ref{sec:sum}
we summarize our results and present our conclusions.


\section{OBSERVATIONS}
\label{sec:obs}

\subsection{Sample Selection}
\label{sec:obs1}

The largest spectral catalog of S stars is Stephenson's second edition
(1984), which includes 1347 objects with poorly known positions.  The
stars in this catalog were originally identified from blue, red, and
infrared objective prism plates.  The majority of the S stars were
discovered on the basis of the red system of ZrO absorption bands,
with a bandhead around 6474\,{\AA}.  Since the Stephenson catalog has
a limiting V magnitude of approximately 11.5, most of the objects are
highly saturated in the SDSS and reliable u, g, r, i, and z magnitudes
have not been found for any known S stars.  We assembled a sample of
known S giants in order to generate likely g and r colors for S giants
and dwarfs.  These constraints on colors of S giants and dwarfs could
allow us to search the SDSS and 2MASS catalogs for potential dwarf S
star candidates. 

To choose the stars making up the sample, we selected 
high Galactic latitude ($|b| > 20$ deg), northern ($\delta > -05 $
deg) S stars from Stephenson's catalog.  This ensures that our sample
is accessible from Mt. Hopkins, and also decreases problems of object
confusion or reddening.   We then correlated the sample from
Stephenson's catalog with 2MASS to find improved positions for most
objects; accuracy of the original positions is typically
$\sim 4\arcsec$, but commonly as poor as 20\arcsec.
The final sample list, which includes 57 objects, can be
found in Table\,\ref{tab:stars}.  Each object is accompanied by its
identification in 
both the Stephenson and 2MASS catalogs, as well as its position.
Since many of these stars are known to be variable, we include the
date of observation.  We include other common identifiers for each
target and a published spectral type, if available.
We include our calculated temperature index for the star, which
is discussed further in \S\ref{sec:temp}.  Spectral types and
temperature indices of the S stars in the sample may vary as a
function of time, because the stars themselves do.  When spectral types for different epochs were available, we
match the $V$ magnitudes of the epochs to those listed in
Table\,\ref{tab:mags} to 
find the spectral type.  Finally, we include identifications of some
stars as intrinsic or extrinsic (denoted respectively by 'i' and 'e')
from \citet{2006AJ....132.1468Y} or from our own calculations based on
AKARI magnitudes.  For more information regarding the
intrinsic/extrinsic identifications, see \S\ref{sec:intext}.

\subsection{Observations and Data Reduction}
\label{sec:obs2}

Spectra were obtained using the FAST instrument on the 1.5 m
Tillinghast reflector on Mt. Hopkins.  FAST produces medium-resolution
optical wavelength spectra, spanning a wavelength range from 3474 to
7418 \,{\AA}.  All data were taken using the 3" slit and a
300\,lines\,mm$^{-1}$ grating, producing a resolution of 
$\lambda / \Delta\lambda\sim$1,000 at 6,000\AA.
Exposure times varied widely based on the magnitude of the star, but ranged from
$\sim$3 to 300 s. The two-dimesional spectra were bias-subtracted and
flat-fielded by the observatory standard data reduction procedure.
After extraction, and the one-dimensional spectra were wavelength
calibrated.  

We used standard star spectra taken each night of the observations to
produce sensitivity curves and flux calibrated spectra.  An extinction
correction file from Kitt Peak National Observatory was also applied
to correct for major atmospheric effects.  The standard calibration
stars used to create the sensitivity curve are BD+26 2606, BD+33 2642,
BD+17 4708, Feige 34, and HD 19445.  Flux calibration and extinction
corrections were performed using the \emph{onedspec} package and the
\emph{calibrate} task in IRAF.  We used a combination of observatory
logs and night-to-night comparison of the sensitivity function to
determine whether the calibrated target spectra were reliable enough
to use in color analysis.  We rejected some of the initial data sample
based on concerns about non-photometric conditions and atmospheric
distortion.  Other nights are included in the final data for analysis
because while the data may not be entirely photometric, it remains
reliable enough for color analysis because of the non-wavelength
dependent nature of most remaining distortions.  We discuss the errors
in our derived SDSS colors further in \S\ref{sec:colors1}. 

Many of the targets are luminous enough in red wavelengths that longer
exposure times cause parts of the spectrum to saturate.  The
wavelengths from 6000\,{\AA} onward are of particular concern.
However, shorter exposure times limited the amount of information we
could derive from wavelengths below 4500\,{\AA} because of low
signal-to-noise ratios.  To create an S star spectral atlas, when
possible we interpolated between two different exposure times:  a
shorter exposure time that produced accurate, unsaturated spectra at
the longer wavelengths, and a longer exposure that resulted in higher
signal-to-noise at shorter wavelengths.  Generally, we determined by
hand which spectra were reliable in a particular wavelength region by
comparing multiple exposure times and examining long exposures for
nonlinear behavior in areas from 6000\,{\AA} onward.  For stars with
only one available exposure time, we rejected a star as saturated if
the counts exceeded 60,000.  After interpolating between spectra to
eliminate saturated regions and  rejecting some data, we were left
with 46 S star spectra.   

\section{Creating the Spectral Atlas}
\label{sec:spatlas}

\begin{figure*}[t]
\begin{center}
\includegraphics[height=4in,width=6in]{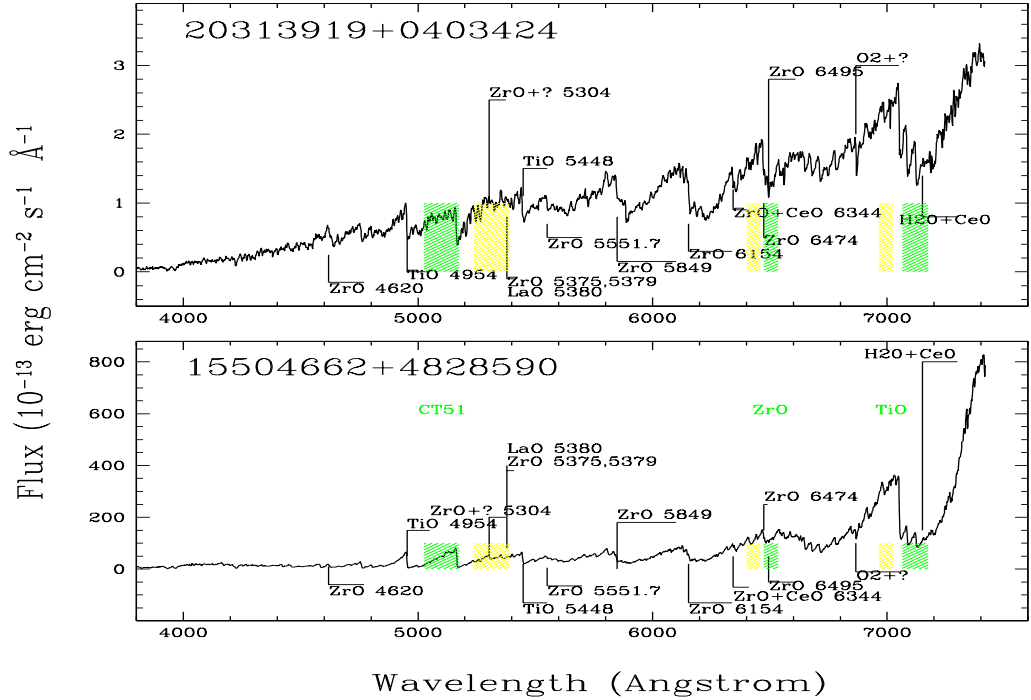}
\caption{ FAST spectra of two different S stars, which are named by
  their 2MASS identifiers.  The bandheads of major absorption features
  are labelled. Flux is in units of $10^{-13}$ \ergscm \AA.  While
  there are few bands unique to one of the spectra, the colors of
  two objects are quite different. The wavelength ranges we use
for spectral indices are shown as colored bars, green within the
molecular absorption band, yellow for the nearby comparison bandpass.
Labeled in green near 7000\AA\, are the TiO index bandpasses, near
6500\AA\, are the ZrO index bandpasses, and near 5200\AA\, are the
bandpasses for the ``CT51'' index, which is intended to measure C$_2$
but may contain substantial effects from TiO.
}
\label{fig:speclines}
\end{center}
\end{figure*}

We use  S star
line identifications \citep{1978MNRAS.184..127W} to identify major
spectral features and select 
objects with unique characteristics to include in the final spectral
atlas, which includes 14 objects marked in Table\,\ref{tab:stars}.
The objects were chosen to be a 
representative sample of the larger FAST sample, which includes many
fundamentally similar objects.  Two sample spectra with associated molecular
line identifications are plotted in Figure\,\ref{fig:speclines}.  While
the spectra display the same molecular bands, most notably those
associated with ZrO, they have different overall colors, which
may be due to differences in abundances and/or effective temperatures.
We also note that these spectra show some TiO bands, indicating that
they are not pure S stars, but instead probably belong to the MS
classification.  The appearance of TiO bands indicates the possibility
of a C/O ratio slightly less than 0.95.  The 
creation of a spectral atlas allows for greater investigation into the
similarities and differences within the S star class.  For instance,
some members of the spectral atlas exhibit, in addition to prominent
ZrO features, H$\alpha$, H$\beta 
$, and H$\gamma$ emission lines.  The presence of these emission lines
indicates that these objects are almost certainly AGB stars,
e.g., Miras.  The
spectral atlas is made available digitally as 14 individual FITS files
via this journal. 

\section{Spectral Indices of S Stars}
\label{sec:specind}

We can immediately use the first digital spectral atlas of S stars
to derive some medium-resolution spectral indices, potentially
useful to classify stars as M, S, or C even where they
may have very similar broadband colors.  Apart from our own FAST spectra
of 48 S stars, we use M giants from the Indo-US spectral atlas
\citep{Valdes04}.  We use eight carbon star spectra also obtained with
the FAST for a separate project (Green et al. 2011, in preparation).
These latter spectra span a wide range of spectral band strengths, but
are selected from the SDSS color wedge defined by \citet{Margon02}.

\begin{figure*}[t]
\center
\includegraphics[width=3.2in,height=3.2in]{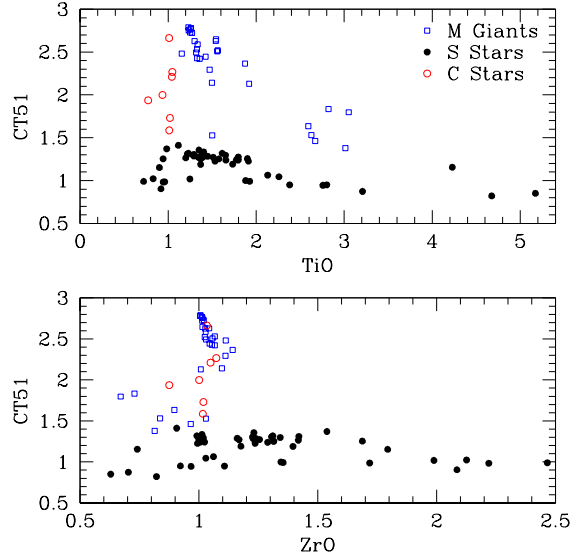}
\caption{Spectral indices of S, M, and C stars.
  ({\it top}) CT51 vs. TiO indices for M giants (open blue boxes),
S stars (filled black circles) and C stars (open red circles).
M and C stars show ZrO indices below $\sim1.2$. 
S stars show a wide range of ZrO indices, but have CT51 indices
less than $\sim 1.4$. }
\endcenter
\label{fig:specind}
\end{figure*}

We defined three indices, one each for TiO, ZrO, and C$_2$.  
We follow the basic premise of using ratios 
of mean flux per \AA ngstr\"om across key molecular bands, so that
spectra of differing resolution should yield similar results.  
We use a neighboring comparison region for each band, so that
sensitivity to broadband flux calibration is minimized. 
For the ZrO index, we divide the mean flux \AA\, across 6400 --
6460\AA\, (ZrO off-band) by that between 6475 -- 6535\AA\, (ZrO band).
For the TiO index, we use mean \AA$^{-1}$ flux from 7065 -- 7175\AA\,
(TiO band) divided by 6965 -- 7028\AA\, (TiO off-band).  The C index
is derived from 5025 -- 5175\,\AA\, (C$_2$ band) and 5238 -- 5390\AA\,
(off-band).  Thus, lower flux in the denominator caused by strong
molecular bands yields a larger spectral index value.
We note that the C index near 5100\AA\, can also be substantially
contaminated by TiO, so we label it ``CT51'' hereafter.
At least within this representative sample of stars, a combination of
these indices yields a preliminary S star classification, as
shown in Figure~\ref{fig:specind}. The objects most reliably
classified as S stars would have ZrO index $>1.2$ and C$_2$ index
$<1.4$.

\section{Color Selection of S Stars}
\label{sec:colors}

\subsection{Generating SDSS and 2MASS Colors}
\label{sec:colors1}

To use the SDSS database to find likely candidates for both S 
giants and dwarfs, we need reliable colors for our objects in as many
SDSS bandpasses as possible.  The spectra from FAST cover the 
bandpasses for the $g$ and $r$ filters completely, and we are thus able to
convolve our spectra with these bandpasses and generate
SDSS colors. Each of the FAST spectra contains 2681 data points, with
approximate wavelength separation of 1.47 \AA.  To
convolve the spectra with the filter transmission curves, we use
linear interpolation between transmission values to match a
transmission constant to each data point in the spectrum.  Once we
find a transmission constant for each wavelength, we normalize the
transmission curve and then convolve the spectrum with the transmission
curve.  We use the IRAF task \emph{calcphot} in the \emph{synphot} package
of \emph{stsdas} to convolve the spectra with the bandpasses and
obtain SDSS colors from our spectra.  The SDSS bandpass curves 
are taken from information made available by the SDSS consortium
\citep{1996AJ....111.1748F}.  As with all SDSS magnitudes, the
resultant magnitudes are in the AB magnitude system.
By including both SDSS and 2MASS bandpasses, we are able to calculate
10 unique colors. 

We also conduct error analysis on the derived SDSS colors using the
standard star spectra and the known g-r colors and r magnitudes for
BD+26 2606, BD+33 2642, BD+28 4211, and BD+174708.  These colors are
published by the SDSS consortium, and allow us to constrain our errors
in both the magnitudes and colors \citep{2002AJ....123.2121S}.  We
calculate the difference between measured and standard $g$ and $r$
magnitudes, and find a standard deviation of  $ \sigma{_g} = 0.191 $
mag for the $g$ band and $ \sigma{_r} = 0.192 $ mag for the $r$ band.
While these seem high and would suggest, via propagation of errors, a
correspondingly high error in our colors (on the order of 0.4 mag), we
find a standard deviation of only $ \sigma{_{color}} = 0.029 $ mag for
the color.  The value was calculated by comparing the derived $g-r$
colors for the standard stars to the known $g-r$ colors from the SDSS
Data Release 5 standard star network \citep{2002AJ....123.2121S}.  
The low value in the uncertainty of our colors is probably due to the
fact that while our spectra may not be absolutely photometrically
calibrated, many of the remaining distortions are not wavelength
dependent and therefore cancel out during color calculation.  Examples
of remaining distortions include slit losses, an incomplete
cancellation of the telluric spectrum or a thin cloud layer that dims
the overall flux at all wavelengths.  So while individual magnitudes
remain unreliable at 
best, the errors in the derived color are comparable to color errors
in 2MASS and are thus reliable enough for our analysis.  The 2MASS
magnitude errors for our sample have a wide range dependent upon the
reliability of the photometry for a particular night, but generally
range from 0.02 magnitudes up to about 0.2 magnitudes.  Colors
calculated from 2MASS magnitudes thus have errors comparable to the
derived SDSS colors.

\subsection{Color Selection Analysis and Criteria}
\label{sec:colors2}

\begin{figure*}[t]
\includegraphics[width=3.2in,height=3.2in]{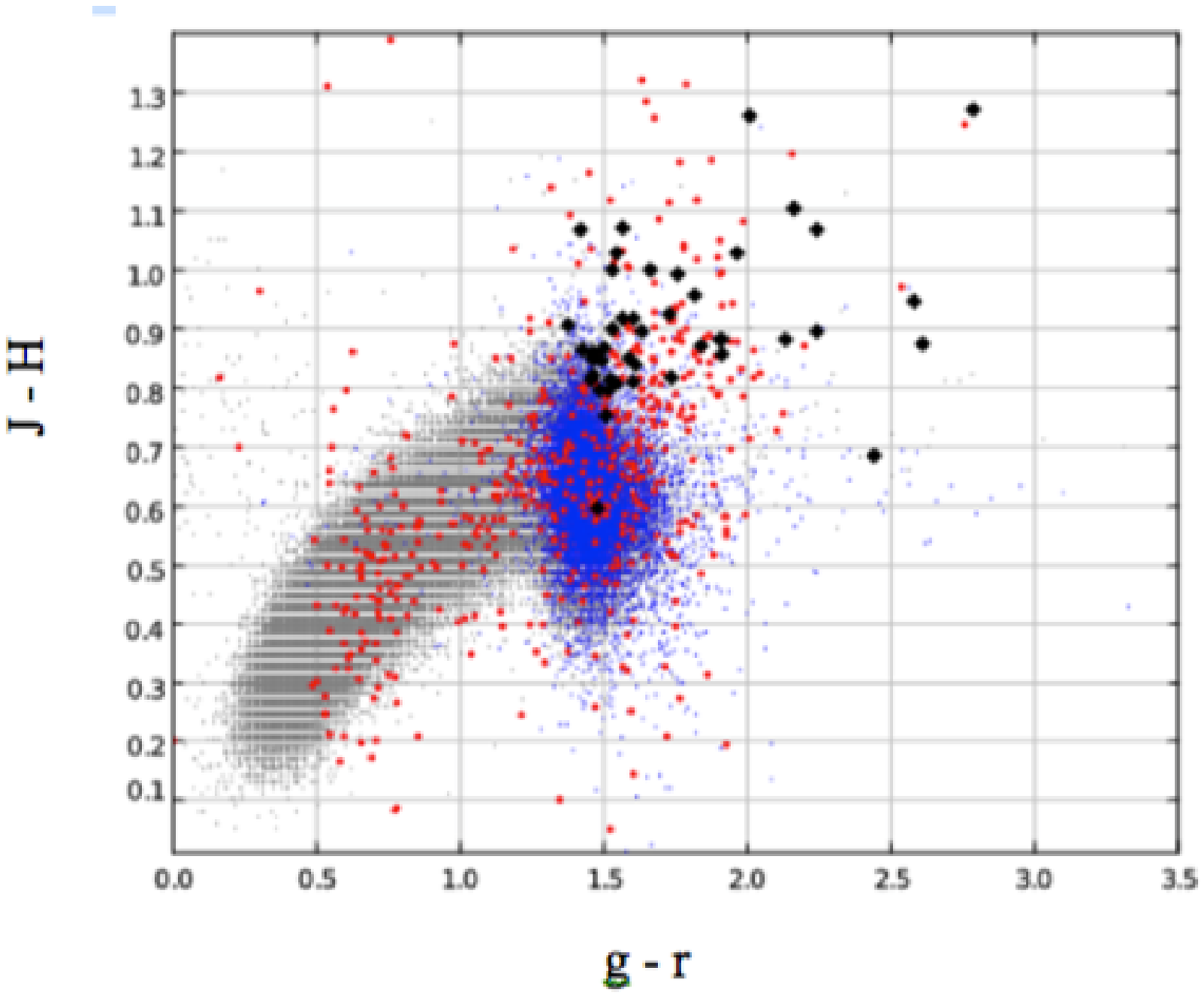}
\hspace{0.3in}
\includegraphics[width=3.2in,height=3.2in]{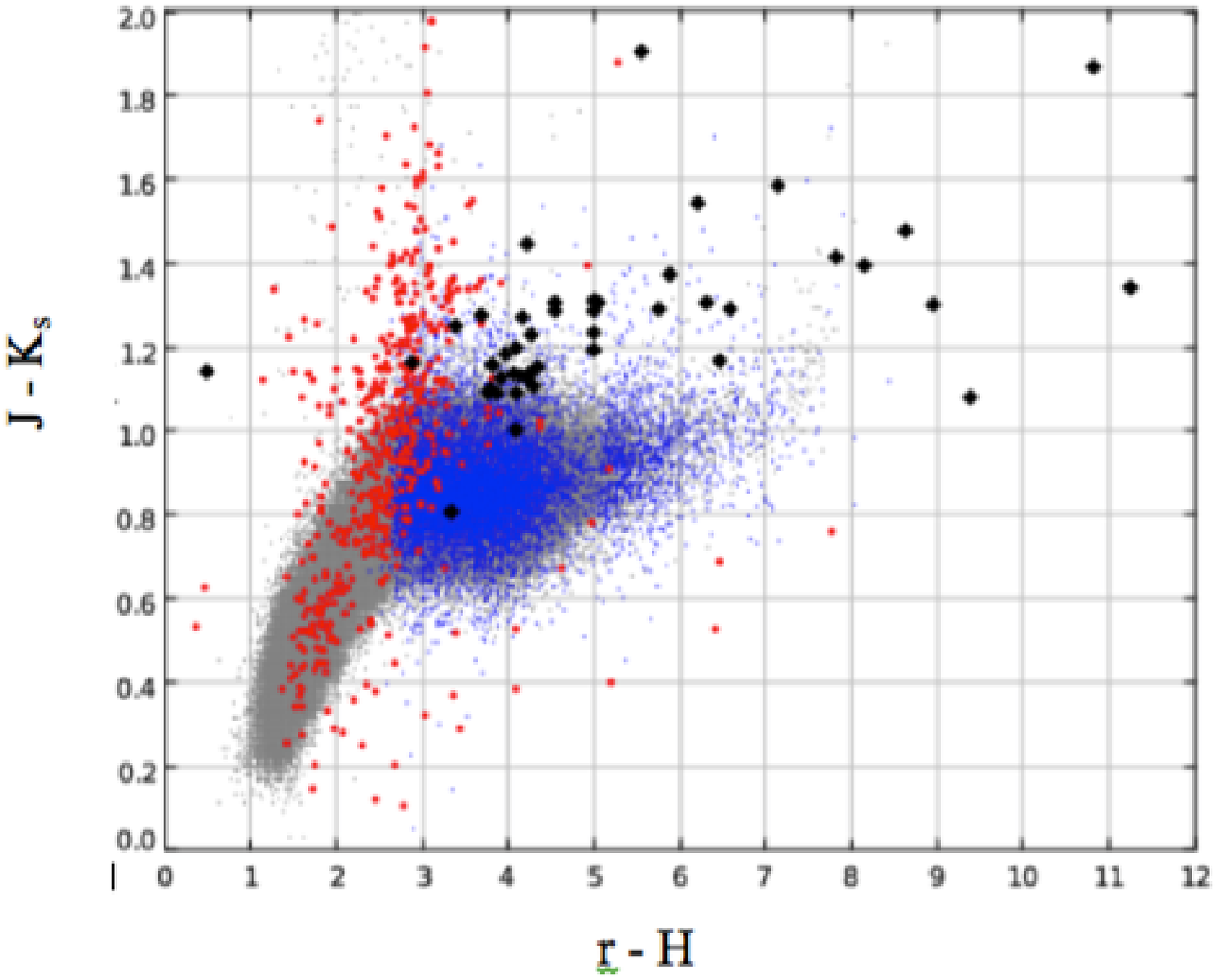}
\caption{Two color-color diagrams, comparing the stellar locus (small
  grey points), the FAST S star sample (black), C stars (red), and
  known M stars(in blue).  Carbon star colors
  are from SDSS and 2MASS  matches, as presented by Green \etal (in
  preparation).  ({\it left})   A plot of $g-r$ color versus $J-H$
  color.  ({\it right}) A plot of   $r-H$ versus $J-K_s$ color.} 
\label{fig:2MASS_SDSScolors}
\end{figure*}

The initial aim of the project was to select for S stars
based on their SDSS and 2MASS colors.  To achieve
this goal, we derive $g$ and $r$ magnitudes for the stars making up
our sample and combine with the known 2MASS magnitudes to construct a
total of 10 unique colors.  The $g$, $r$, and 2MASS magnitudes, along
with $V$ band, IRAS, and AKARI photometry where available, are
presented in Table\,\ref{tab:mags}.  Many of these colors, however,
are ultimately unhelpful for analysis purposes because the colors of
known S stars cover several 
magnitudes.  Such a wide spread in magnitudes does not allow for
efficient discovery of the S stars based on color.  In other color
combinations, the S stars are not well distinguished from the stellar
locus, so selection based on color is contaminated by large numbers of
stars from the main sequence.  We find that the most useful color
combinations are $g-r$, $J-H$, and $H-K_s$ because in these colors,
the FAST S stars sample is well distinguished from the stellar locus.
We compare these colors with those found for general SDSS stars,
by overplotting some 300,000 SDSS/2MASS matches of
\citep{2007AJ....134.2398C}, which they used to parametrize the stellar
locus in terms of color.  In Figure\,\ref{fig:2MASS_SDSScolors}, we
present two color-color diagrams in which the S giants from the FAST
sample are best separated from the stellar locus.  The first shows the
comparison of  $g-r$ color and $J-H$ color, while the second shows the
comparison $r-H$ and $J-K_s$ color. 

However, the S giants that comprise the FAST sample, by virtue of
their location on the TP-AGB or RGB, have similar colors and effective
temperatures to M giants.   In  Figure\,\ref{fig:2MASS_SDSScolors}, we
also examine colors for M giants and dwarfs  matched in the SDSS and
2MASS surveys.  While the M giants 
are very few in number (due to the depth of both of these surveys,
many nearby M giant targets are saturated), we can immediately see
that S stars and M stars lie in essentially the same area of color
space.   We also include C stars on the color-color diagrams, since
they follow evolutionary paths similar to S stars and also have 
enhanced abundances.  The S stars of the FAST sample, while
well-distinguished from the overall stellar locus on both color-color
diagrams, are particularly difficult to distinguish from C stars on
the $g-r$ versus $J-H$ diagram.  
On the plot comparing $r-H$ to $J-K_s$, S stars
can be  distinguished reasonably well from the bulk of the M star
population using $J-K_s > 1.2$ and $r-H > 3.5$.  Given the biases in
the current sample of S stars, it is difficult to determine the
efficiency of this selection, but this region on the color-color plot
appears to be the least contaminated by M and C stars.   
Spectroscopy for a sample of perhaps 100 stars in this
color region could directly test the efficiency of this color selection.

\section{Distinguishing Intrinsic and Extrinsic S Stars}
\label{sec:intext}

\subsection{Techniques}
\label{sec:intext1}

Distinguishing intrinsic and extrinsic S stars is difficult without
high-resolution spectroscopy, where direct detection of Tc is
possible.  However, as presented in Van Eck and Jorrissen (2000),
there are other measurements which serve to
segregate the two types with reasonable statistical accuracy.
Since extrinsic S stars are generally on the RGB (or early
AGB), they cluster closer to the main sequence than intrinsic S stars
in terms of color.  Stars on the AGB generally show a red excess which
is absent on the RGB.  However, Van Eck and Jorissen (2000) find that
blue excursions of Mira variables, which make up a significant
proportion of intrinsic S stars, make the two classes difficult to
distinguish based on distance from the main sequence alone.
Additionally, intrinsic and extrinsic S stars have been shown to
segregate well in a ($K-[12]$, $K-[25]$) diagram
\citep{{1993A&A...271..180G},{1993A&A...271..463J}}.  These colors are
based on IRAS photometry, which is available for many of the stars in
the FAST sample.  IRAS magnitudes in these bands are presented in
Table\,\ref{tab:mags} for 37 stars (with two bands available for 29).
Identifications of many of the S stars in the 
FAST sample as intrinsic or extrinsic are included in
Table\,\ref{tab:stars}.  All such 
identifications not marked with an asterisk are from
\citet{2006AJ....132.1468Y}.  Classification of most of the S stars
in our sample is inconclusive because of their intermediate positions
in the color-color diagrams.  However, this IR technique
remains the most efficient photometric classification method 
known to determine whether S stars are intrinsic or extrinsic. 

We also use results from Vanture \etal (private communication), which
indicate that S stars presenting Li in their spectra are intrinsic S
stars.  Lithium in stars is destroyed at fairly low temperatures
\citep{1965ApJ...142..451B} but can be produced by the Cameron-Fowler
mechanism, in high luminosity ($M_{Bol}<-6$) TP-AGB stars via hot
bottom burning. Since, as mentioned above, extrinsic S stars are often
less evolved along the giant branch than intrinsic S stars, they
generally lack any evidence for the resonance line Li 6707\AA\,
(Vanture et al. 2011,  in preparation).  We compare the list of S stars with
and without Li detections to our FAST sample and analyze them for the
Li 6707\,{\AA}\, feature.  Unfortunately, we find that this line
cannot be reliably detected with spectra of our resolution and S/N.

\subsection{Identification Using AKARI Magnitudes}
\label{sec:intext2}

\begin{figure*}[t]
\center
\includegraphics[width=2.7in,height=2.7in]{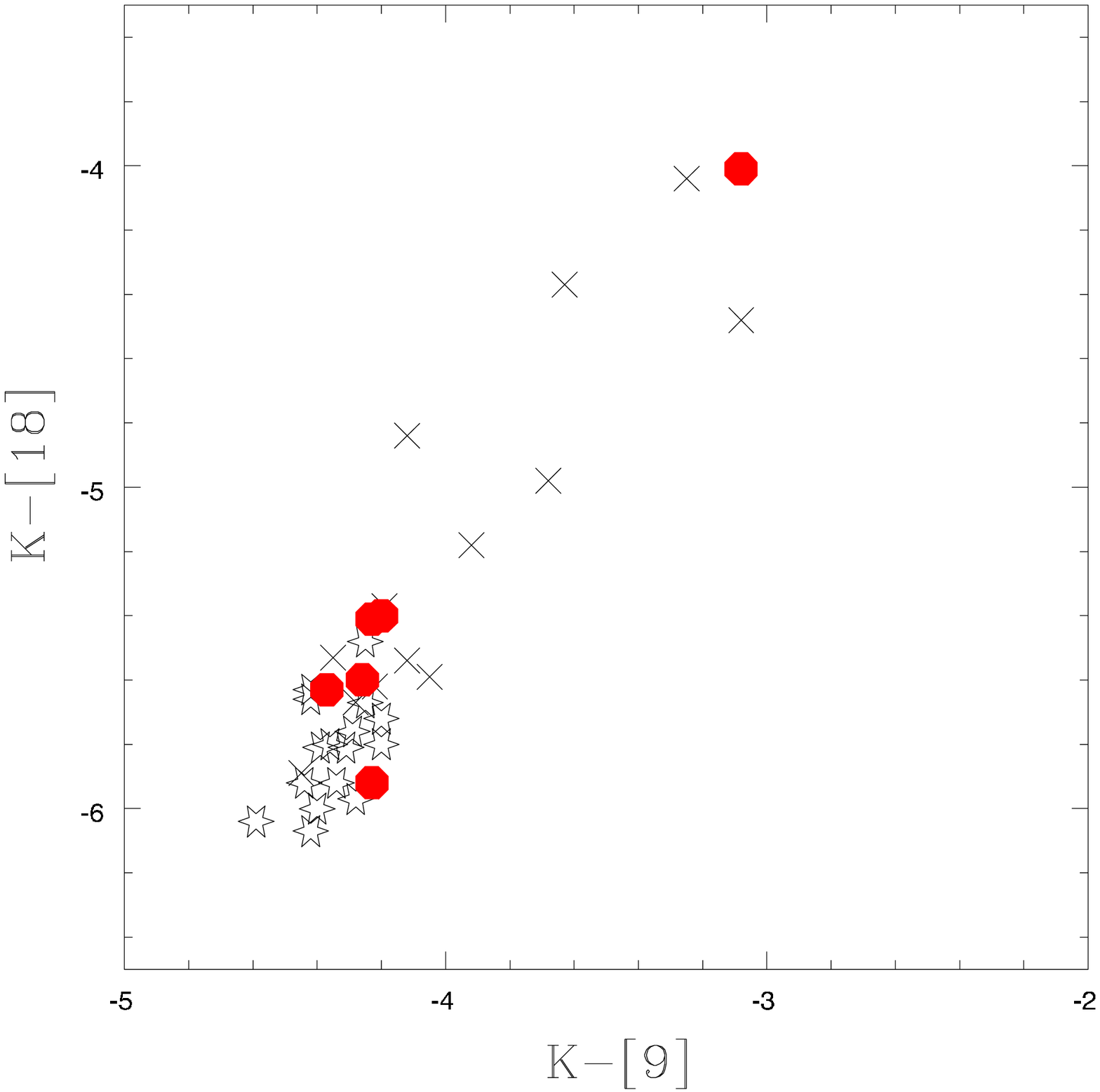}
\hspace{0.3in}
\includegraphics[width=2.7in,height=2.7in]{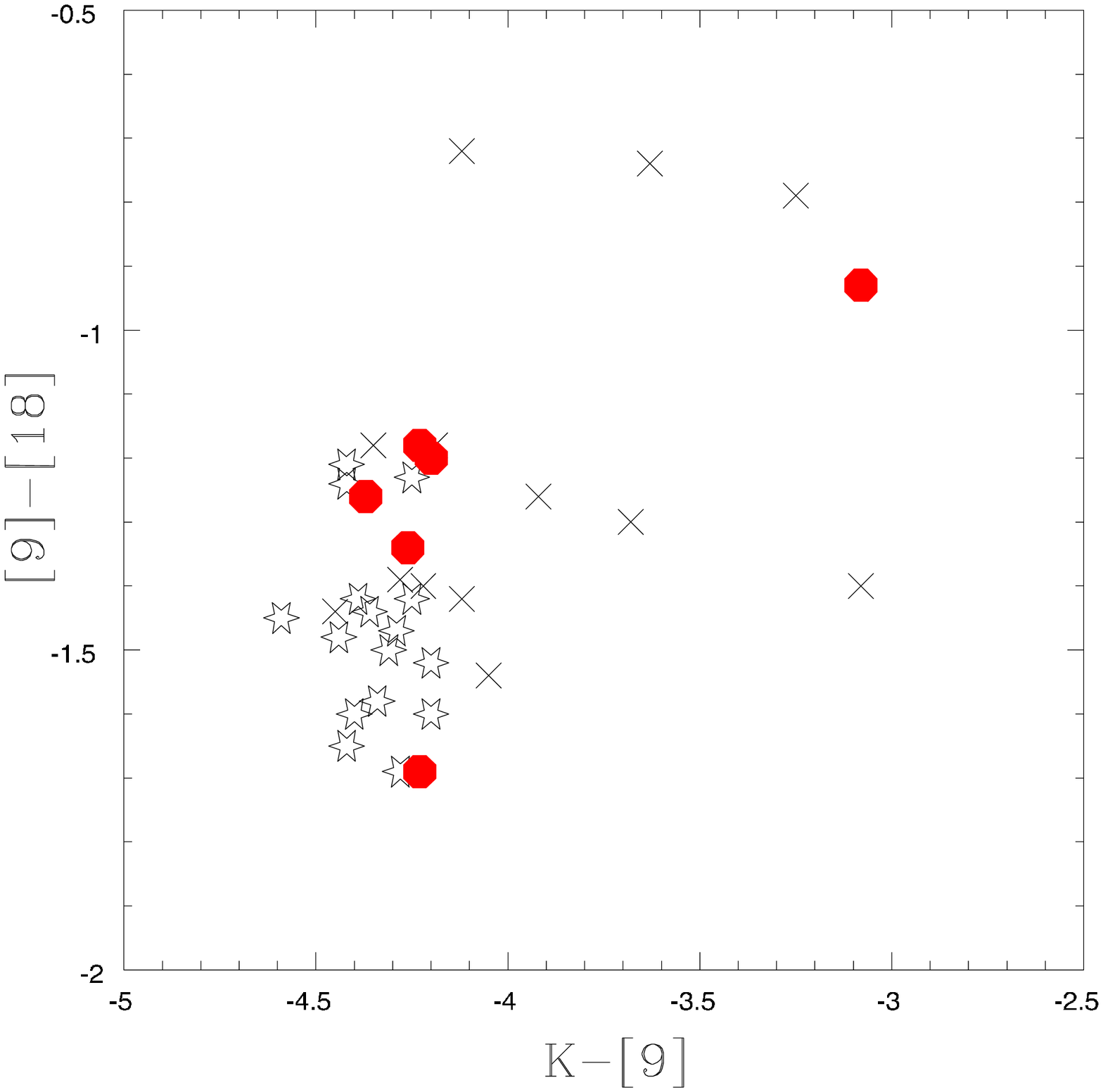}
\caption{A comparison of 2MASS $K$ magnitudes and AKARI infrared
  magnitudes to further distinguish intrinsic and extrinsic S stars.
  Known intrinsic stars are plotted as crosses, while known extrinsic
  stars are plotted with star symbols.  Previously unclassified
  objects are plotted in red.  These plots show distinctions between
  intrinsic and extrinsic S stars similar to those in Figure 4, which
  uses IRAS magnitudes.  ({\it top left})  A plot comparing $K-[9]$
  and $K-[18]$ color.  ({\it top right})  A plot comparing $K-[9]$ and
  $[9]-[18]$ color.} 
\endcenter
\label{fig:nearMidIRcolors}
\end{figure*}

The AKARI satellite surveyed much of the sky in both near and far
infrared bands. We use flux data collected by  AKARI in
the S9W and L18W bandpasses to find infrared magnitudes for six
objects in the FAST sample without available IRAS colors
\citep{{2007PASJ...59S.369M},{2010A&A...514A...1I}}.  
The S9W and L18W bandpasses are qualitatively similar to the
IRAS 12 and 25 micron bandpasses.  All available AKARI magnitudes are
presented in Table\,\ref{tab:mags}, for 52 stars (with two bands
available for 35). Since the zero magnitude fluxes for these
bandpasses are not well established, all magnitudes are presented on
the AB system.  In Figure\,\ref{fig:nearMidIRcolors}, we show two
color-color diagrams comparing intrinsic and extrinsic stars using
$K-[9]$, $K-[18]$, and $[9]-[18]$ colors.  S stars 
previously classified as intrinsic vs. extrinsic 
are shown in black, with others plotted in red. 

From the AKARI magnitudes and plots, we are able to identify an
additional two stars in the sample as intrinsic or extrinsic.  We
identify 22315839+0201206 as an extrinsic S star.  This star has a
$K-[18]$ value of $-5.92$ mag and a $[K]-[9]$ value of $-4.23$ mag.
This places it in the extreme lower left corner of the color-color
diagram comparing $K-[9]$ color to $K-[18]$ color, far away from any
known intrinsic stars but close to many known extrinsic stars.
Similarly, we are able to identify 19364937+5011597 as an intrinsic S
star on the basis of color.  This star has a $K-[9]$ value of only
$-3.09$ mag, and a $K-[18]$ value of $-4.01$ mag.  This places it in
the upper right-hand corner of the plot comparing $K-[9]$ and and
$K-[18]$ color, far away from the cluster of extrinsic S stars.  The
four other stars for which we have no classification
(03505704+0654325, 10505517+0429583, 13211873+4359136, and
16370314+0722207) cluster close to the transition between intrinsic
and extrinsic S stars in all three color comparisons, meaning that we
cannot confidently identify these with either class of stars.  We
include our additional identifications in Table\,\ref{tab:stars},
marked with an asterisk to distinguish them from those from
\citet{2006AJ....132.1468Y}. 

\section{Calculating Temperature Indices}
\label{sec:temp}

\subsection{Using Colors to Determine Temperature}
\label{sec:temp1}

As mentioned before, S stars cover a wide range of spectral types and
effective temperatures.  However, they generally have effective
temperatures similar to those of M giants.  The starting point for
classifying the spectral type and temperatures of our S stars,
therefore, is to calculate a temperature index using M-giant criteria
\citep{Gray09}. \citet{2000AJ....119.1424H} use a grid of
stellar models to calibrate a relation between color and temperature
index for M stars.  Their analysis uses CIT/CTIO colors $V-K$ and
$J-K$.  We find $V$ magnitudes for some of the FAST sample using
Simbad.  We also convert the 2MASS K$_s$ magnitudes and $J-K_s$ colors
using the following relations from
\citet{2001AJ....121.2851C}:\footnote{Updated at
  http://www.astro.caltech.edu/\~jmc/2mass/v3/transformations}    
\begin{align}
&K_s = K_{CIT} + (-0.019\pm0.004) + (0.001\pm0.005)(J-K)_{CIT} \notag \\
&J-K_s = (1.068\pm0.009)(J-K)_{CIT} + (-0.020\pm0.007)
\end{align} 
We then use the temperature index/color relation to assign a
preliminary temperature index to the stars in the FAST sample.  Where
possible, we use the $(V-K)_{CIT}$ relation, since this is specified
by the authors as the most temperature sensitive.  When there is no
$V$ magnitude available, we use the $(J-K)_{CIT}$ relation
\citep{2000AJ....119.1424H}.   

\subsection{Results of Temperature Analysis}
\label{sec:temp2}

Results of this analysis are presented in Table\,\ref{tab:stars}.
These temperature indices should be treated with caution, since many
of the stars in the FAST sample are highly variable.  
Furthermore, differences in C/O ratios, may lead to temperature errors
up to 400\,K  \citep{2010arXiv1011.2092V}.  However, the preliminary 
classification gives us a rough idea of the relative effective
temperatures of the S stars.  We find that the average temperature
index of the stars in the sample is 5, which in M giants corresponds
to an effective temperature of roughly 3500 K
\citep{2000AJ....119.1424H}.  In Table\,\ref{tab:stars}, we also
present known spectral types from \citet{1980ApJS...43..379K}, which include
temperature indices for the S stars.  We find reasonably good
agreement between our calculated temperature indices and those
presented as part of the spectral type.  Differences between
our temperature index and those included as part of the published
spectral types are within the range of variability of the star.  We
also note that since we do not have a reliable color/temperature
relation for stars with a temperature index greater than 7, we present
all of these classes as 7+.  We also find reasonably good agreement
(within one temperature index) between the values using $(V-K)_{CIT}$
and those derived using $(J-K)_{CIT}$. 

We also conduct a preliminary analysis on possible correlations
between the intrinsic/extrinsic distinction and the temperature
indices.  We find that the average temperature index of the intrinsic
stars is $5.6\pm2.3$, while the average temperature index of extrinsic
stars is $4.3\pm2.2$.  Therefore, while the mean temperature index
shows some differences, this measurement alone is not statistically
significant.  A larger sample would be needed to determine the
validity of differences between average effective temperature between
intrinsic and extrinsic S stars.  The temperature indices of S stars
in general are insufficient to classify them as intrinsic or
extrinsic, except for the reddest objects.

\section{Parallax, Proper Motion, and Absolute Magnitude}
\label{sec:motion}

\subsection{Parallax Analysis}
\label{sec:motion1}

Four of the stars in the original FAST sample have reliable parallax
measurements (defined as a parallax detection at the $3\,\sigma$
level) available from the Hipparcos survey
\citep{1997A&A...323L..49P}, so calculation of the 
absolute magnitude of these stars in the $g$ and $r$ bands is
possible.   We calculate the error in our absolute 
magnitudes using the presented error in the parallax measurements and
assuming apparent magnitude error values of $ \sigma{_g} = 0.191 $ and
$ \sigma{_r} = 0.192 $.  Results of this analysis are given in 
Table\,\ref{tab:plx}.  We note the calculated absolute $g$ and $r$
magnitudes, as well as 
the associated errors.  Objects are identified using the 2MASS
Identifier, and the parallax and associated error as given by the
Hipparcos catalog are also included. We can immediately identify each
of these four objects as asympotic or red giant branch members:  if
they were dwarf stars of the same spectral type, we would expect
magnitudes in the $g$ and $r$ of +5 or higher, as opposed to the
derived values, which center around 0. 

\subsection{Proper Motion Analysis}
\label{sec:motion2}

We also find that some stars in the original FAST sample display
significant (again, $3\,\sigma$) proper motions. The proper motions
are available in the Tycho-2 catalogue \citep{1997A&A...323L..57H}. Of
these, we have reliable $g$ and $r$ photometry for 16 objects.  Given
the proper motion and assuming that all objects are moving at total
space velocities less than the Galactic escape velocity, we can derive
an upper limit on the distance, and thus also constrain the absolute
magnitudes of the objects.  To derive an approximate distance ($d$),
we use the equation: 
\begin{align}
d &= \frac{v_{trans}}{\mu} < \frac{v_{esc}}{\mu} \\
d &< 0.0056\,pc/yr \left(\frac{\mu}{1\,rad/yr}\right)^{-1} \\
d &< 1.16\,kpc/yr \left(\frac{\mu}{1\,mas/yr}\right)^{-1}
\end{align}  
where $\mu$ is the proper motion of the object measured in radians per
year, $v_{trans}$ is the transverse velocity of the object, and $v_{esc}$
is the Galactic escape velocity.  We take the Galactic escape velocity
to be 545 km/s \citep{2007MNRAS.379..755S}, which is equivalent to
0.0056 parsecs per year.  The distance constraint is only an upper 
limit because we neglect radial velocities.  However, we note that our
{\tt rvsao} measurements generally yield small radial velocities
($\lax$50\,km/s) for the S star sample.  We then use the same formulas
presented above to derive 
$g$ and $r$ absolute magnitudes and associated errors.  These $g$ and
$r$ absolute magnitudes essentially represent the brightest the object
concerned could be, considering its proper motion and assuming that
all observed objects are gravitationally bound members of the Milky
Way. Results of the analysis are presented in Table\,\ref{tab:pm}.  We can
immediately see from these lower bound magnitudes that these objects
could be RGB or AGB stars - even considering that the actual
magnitudes are probably significantly fainter.  For the sample of 16
stars, we find that the average lower limit on the absolute $g$
magnitude $M_g$ is $-5.0$ mag, correspondingly $M_r \sim -6.7$ mag.
Cool dwarf stars generally have $g$ and $r$ absolute magnitudes
greater than +8 mag
\citep{{1977ApJ...216..757D},{1947PMcCO...9..197V}}  We conclude that
all 16 objects displaying significant proper motions could be giants. 

\subsection{Lower Limit Distances}
\label{sec:motion3}

To further bear out this analysis, we also determine a lower limit on
the distance for these objects by assuming (conversely to above) that
they {\em do} have a dwarf magnitude.  We take the approximate absolute
$g$-band magnitude for an M dwarf star to be $M_g$ = 10 mag and the
absolute $r$-band magnitude to be $M_r$ = 9
mag \citep{{1977ApJ...216..757D},{1947PMcCO...9..197V}}.  This method
can be applied to the entire sample, because it does not rely on
proper motion or parallax measurements.  We find that most of these
objects, were they to have typical dwarf magnitudes, would be within
25 pc of us.  This distance would almost guarantee that these objects
would have detectable parallax and proper motion.  For instance,
assuming extrinsic S stars are a mostly spheroid population (like dC
or CH stars:  see \citet{1994ApJ...434..319G} and
\citet{Bergeat02}) with velocity of $-220$\,km/s relative to
the Sun \citep{2004AJ....127..914S}, the typical proper motion for a
star 25 pc away would be on the order of 2.25 mas/yr.  Additionally,
such stars would have a parallax on the order of 40 mas. Since such a
parallax would be easily detectable by Hipparcos, these lower limit
distances therefore also suggest that the stars of the FAST sample are
likely to be giants or AGB stars.

\section{Summary and Conclusions}
\label{sec:sum}

Using a sample of 57 medium-resolution S star spectra taken with the
FAST spectrograph, we create a spectral atlas of S stars comprised of
14 objects that span a range of spectral types within the MS, S, and
SC classes.  The atlas is published as a collection of 1-D FITS
files via this journal.  After generating $g$
and $r$ SDSS magnitudes from the spectra, we find that the SDSS
magnitudes in the $g$ and $r$ bands, combined with $J$, $H$, and $K_s$
magnitudes from the 2MASS catalog for S stars may with further
confirmation allow for reasonably efficient color selection of
these objects from the SDSS and 2MASS catalogs.    We use
previously published data to identify some of the stars in the sample
as intrinsic or extrinsic stars, and find that the fraction of
extrinsic S stars in the sample is approximately $54\%$.  We also
assign temperature indices to the stars in the sample based on the M
star scale of temperature indices.  We find that much of the S star
sample falls at temperature index 4 or above, meaning that the
effective temperatures of most of these stars are well below 5000 K.
We also analyze objects in the FAST sample with detectable parallaxes
and proper motions to generate absolute magnitude limits, as
well as lower limit distances based on assumed dwarf magnitudes.  This
analysis bears out our initial assumption that  the FAST sample is
primarily composed of giant stars, either on the AGB or RGB.

\section{Acknowledgements}
\label{sec:ack}

Many thanks to the referee for a thorough reading.
We are grateful to Kevin Covey for the use of the high-quality sample
of SDSS/2MASS matches. Also, thanks to Warren
Brown and Mukremin Kilic, who provided many helpful discussions and
hints on color selection and the SDSS photometry.  Many thanks to Doug
Mink for all his invaluable help with \emph{rvsao} and SDSS
correlations, and to Bill Wyatt for his help with accessing SDSS DR-7
spectra, all of which we hope to use in upcoming publications.   We
gratefully acknowledge Bob Kurucz, Andrew Vanture and 
George Wallerstein for illuminating discussions.  Thanks also to Verne
Smith and GW for compiling an initial list of Li 6707\,\AA\,detections
in S stars.  I am grateful to the SAO REU program organizers - Marie
Machacek, Christine Jones, and Jonathan McDowell - for all their
support.  Finally,  This work is supported in part by the National
Science Foundation Research Experiences for Undergraduates (REU) and
Department of Defense Awards to Stimulate and Support Undergraduate
Research Experiences (ASSURE) programs under Grant no. 0754568 and by
the Smithsonian Institution.


\clearpage
\begin{deluxetable}{cccclclcc}
\tabletypesize{\scriptsize}
\setlength{\tabcolsep}{0.02in}
\tablewidth{0pt}
\tablecaption{Target S Stars\label{tab:stars}}
\tablehead{
\colhead{Stephenson}&\colhead{RA} &\colhead{Dec} & \colhead{2MASS Identifier}& 
\colhead{Other Identifiers} & \colhead{Obs. Date} & \colhead{Spectral Type} &\colhead{Temp.} &\colhead{Intrinsic}\\
Catalog &  J2000 & J2000 & & & & &Class & vs. Extrinsic \\
Number &(deg) &(deg) & & & & & & }
\startdata
9&   006.008240& +38.577049& 00240197+3834373$^A$&	HD 1967/R And&09/15/09	 &S5e Zr5 Ti2 	 &  	7& i \\
22&  016.300978& +19.197842& 01051223+1911522&	HD 6409/CR Psc&	09/15/09&	 	 &	4/5 & e \\
32&  021.413226& +21.396074& 01253917+2123458&	RX Psc&	09/15/09 	 &	 	 &	0/1 & i\\
45&  028.582236& +21.889090& 01541973+2153207&	BD+21 255&	09/15/09&S3 Zr1 Ti3	 &	 	3/4 & e\\
57&  036.476487& +38.122776& 02255435+3807219&	BI And&	09/15/09	 &S8 Zr7 Ti4	 	 &	7 & i\\
73&  052.040388& +17.679628& 03280969+1740466&	\nodata&	09/20/09&	 &	 	4/5 & \nodata \\
74&  052.936921& +04.695488& 03314486+0441437$^A$&	\nodata&	09/20/09&	 &	 	6 &\nodata \\
80&  055.889249& +22.437023& 03433341+2226132&	BD+21 509&	09/16/09&	 &	 	5/6 & e\\
83&  057.737708& +06.909048& 03505704+0654325&	\nodata&	09/16/09&	 &	 	6 &\nodata \\
94&  066.090802& -02.532759& 04242179-0231579&	BD-02 891&	09/20/09&S2 Zr2- Ti2	 &	 	4 & e\\
106& 073.937675& +79.999931& 04554504+7959597&	BD+79 156&	09/20/09&S4 Zr2- Ti3-	 &	 	4/5 & e\\
134& 080.836187& -04.570626& 05232068-0434142&	HD 35273/V535 Ori&09/20/09&	 	 &	7+ & i \\
312& 108.717017& +68.804321& 07145208+6848155$^A$&	HD 54587/AA Cam&11/10/09&M5S$\dagger$ &	 	4/5 & i\\
339& 111.489470& +62.591026& 07255747+6235276$^A$&	\nodata&	11/10/09&	 &	 	4 &\nodata  \\
347& 112.048398& +45.990597& 07281161+4559261&	HD 58521/Y Lyn&	11/10/09&M6S$\dagger$	 &	 	7+ & i\\
403& 117.325749& +23.734451& 07491817+2344040&	HD 63334/T Gem&	11/10/09&S3e Zr2.5 Ti2	 &	 	4/5 & i\\
405& 117.681704& +47.003967& 07504360+4700142&	\nodata&	11/10/09&	 	 &	4/5 & \nodata \\
413& 118.221777& +34.614128& 07525322+3436508&	BD+34 1698&	11/10/09&	 	 &	6 & i\\
418& 118.369741& +17.780506& 07532873+1746498&	HD 64209&	11/10/09&	 	 &	4/5 & e \\
431& 119.233374& +31.167286& 07565600+3110022$^A$&	AO Gem&	11/10/09	 &	 	 &	7  & i\\
460& 121.836937& +11.321875& 08072086+1119187&	\nodata&	03/10/10&	 	 &	4/5 & \nodata \\
471& 122.761565& +08.139297& 08110277+0808214&	\nodata&	03/10/10&	 	 &	5/6 & e \\
494& 125.428453& +17.285120& 08214282+1717064&	HD 70276/V Cnc&	03/10/10&S0.5e Zr0 &	 	1 & i \\
589& 137.661696& +30.963114& 09103880+3057472&	HD 78712/RS Cnc&03/10/10&M6S$\ddagger$	 	 &	7+ & i\\
612& 143.901826& +69.157043& 09353643+6909253&	BD+69 524&	03/11/10&	 	 &	3/4 & e\\
707& 162.729912& +04.499532& 10505517+0429583&	\nodata&	03/11/10&	 	 &	6 & \nodata \\
722& 166.970132& +68.366440& 11075283+6821591&	HD 96360/HL UMa&03/11/10&	 	 &	4/5 & e \\
803& 190.986134& +61.093254& 12435667+6105357&	HD 110813/S UMa&03/11/10&S6e Zr6 &	 	5/6 & i\\
819& 200.328046& +43.987118& 13211873+4359136&	AV CVn&	03/11/10	 &S3 Zr2 Ti1 &	 	3/4 &  \nodata \\
833& 207.141756& +31.999107& 13483402+3159567&	\nodata&	03/11/10&	 	 &	0 & \nodata \\
836& 208.005660& -03.480082& 13520135-0328482&	HD 120832&	03/11/10&	 	 &	4/5 & e\\
856& 214.462975& +83.831696& 14175111+8349541$^A$&	HD 1272266/R Cam&06/24/09&S2e Zr2 	 &	3 & e\\
855& 216.950026& -03.079711& 14274800-0304469&	BD-02 3848&	05/03/09&	 	 &	4 & e\\
892& 232.384097& +00.189076& 15293218+0011206&	\nodata&	05/03/09&	 	 &	2/3 & \nodata  \\
902& 238.100360& -23.352097& 15522408-2321075&	\nodata&	05/03/09&	 	 &	6 & e\\
903& 237.694254& +48.483078& 15504662+4828590$^A$&	HD 142143/ST Her&05/03/09&M6.5S Zr1 Ti6+ &	 	5/6 & i\\
926& 245.401029& +56.877048& 16213624+5652373$^A$&	HD 147923&	05/03/09&	 	 &	2 & e\\
932& 249.263106& +07.372439& 16370314+0722207$^A$&	BD+07 3210&	05/03/09&	 	 &	4/5 & \nodata \\
958& 256.694247& -02.017769& 17064661-0201039&	\nodata&	05/03/09&	 	 &	6 & \nodata \\
981& 260.529640& +23.817570& 17220711+2349032&	BD+23 3093&	05/03/09&S3.5 Zr3+ Ti4 &	 	4/5 & e\\
986& 261.200788& +33.918953& 17244818+3355082&	\nodata&	05/03/09&	 	 &	3 & \nodata \\
1002&267.252799& +21.710087& 17490067+2142363$^A$&	\nodata&	05/03/09&	 	 &	4 & e\\
1065&279.398924& +43.256588& 18373574+4315237&	\nodata&	05/03/09&	 	 &	3 & e\\
1087&282.875526& +48.911934& 18513012+4854429$^A$&	TU Dra&	05/03/09	 &	 	 &	3/4 & i\\
1150&294.205739& +50.199917& 19364937+5011597&	HD 185456/R Cyg&05/03/09&S5e Zr5 Ti0 &	 	7 & i* \\
1152&294.345155& +67.188873& 19372283+6711199&	\nodata&	05/03/09&S5- Zr4.5 &	 	5/6 & i\\
1165&297.641369& +32.914139& 19503392+3254509$^A$&	HD 187796/chi Cyg&05/03/09&S6+e Zr2 Ti6.5 &	 	7+ & i\\
1222&307.913302& +04.061802& 20313919+0403424$^A$&	\nodata&	05/23/09&	 	 &	5 & \nodata \\
1247&313.355676& +06.539369& 20532536+0632217&	\nodata&	05/23/09&	 	 &	2/3 & \nodata \\
1262&317.601674& +01.606119& 21102440+0136220&	\nodata&	05/27/09&	 	 &	4/5 & \nodata \\
1264&318.232437& +09.548880& 21125578+0932559&	\nodata&	07/17/09&	 	 &	3/4 & \nodata \\
1290&331.995242& +29.665623& 22075885+2939562&	\nodata&	07/17/09&	 	 &	4/5 & e\\
1299&337.993328& +02.022399& 22315839+0201206&	\nodata&	07/17/09&	 	 &	5 & e*\\
1304&340.954936& +33.926231& 22434918+3355344&	HD 215336&	07/17/09&	 	 &	1/2 & e\\
1315&343.648463& +16.941910& 22543563+1656308&	HR Peg&	07/19/09	 &S4+ Zr1.5 Ti4 & 	4/5 & i\\
1328&349.035110& +28.863674& 23160842+2851492$^A$&	\nodata&	07/19/09&	 	 &	3 & e\\
1340&356.540618& +34.783325& 23460974+3446599&	\nodata&	07/19/09&	 	 &	3 & \nodata \\
\enddata
\tablecomments{The 58 S stars targeted by the FAST survey.
The Stephenson Catalog number corresponds to the object's
identification in The General Catalog of S Stars, second edition
\citep{1984PW&SO...3....1S}.  The 14 stars included in our digital
spectral atlas are marked by an $A$ in the 2MASS Identifier. Targets
were initially selected from this survey and then matched with 2MASS
counterparts \citep{2006AJ....131.1163S} to provide the J, H, and K$_s$
magnitudes. Spectral types are taken from \citet{1980ApJS...43..379K}
wherever possible. In some cases, they show types for multiple
epochs, in which case we chose the epoch whose listed $V$ mag in Skiff
\etal (2010) is closest to the $V$ mag we list in Table 2.  If those
types are not available, we list types from
\citet{1954ApJ...120..484K} as $\dagger$ and
\citet{1979ApJ...234..538A} as $\ddagger$. 
We also present our own calculations of the effective temperature,
here presented on the M giant scale, as described in \S\ref{sec:temp1}.
Intrinsic/extrinsic identifications denoted with an asterisk are
derived using AKARI, rather than IRAS, photometry.  For more
discussion, see \S\ref{sec:intext2}.}  
\end{deluxetable}

\clearpage
\begin{deluxetable}{crrrrrrrrrr}
\tabletypesize{\scriptsize}
\setlength{\tabcolsep}{0.02in}
\tablewidth{0pt}
\tablecaption{S Star Magnitudes\label{tab:mags}}
\tablehead{
\colhead{2MASS Identifier}&\colhead{$g$}&\colhead{$r$}&\colhead{$V$}&\colhead{$J$}&\colhead{$H$}&\colhead{$K_s$}&\colhead{$[S9W]$ AB}&\colhead{$[12]$}&\colhead{$[L18W]$ AB} &\colhead{$[25]$}}
\startdata
00240197+3834373&8.40&6.37&7.39&2.02&0.77&0.12&\nodata&$-2.66$&3.17&$-2.3$ \\
01051223+1911522&9.38&7.79&\nodata&3.67&2.76&2.48&6.68&1.75&8.28&1.63 \\
01253917+2123458&16.53&14.40&8.80&6.27&5.39&4.98&8.90&3.91&10.16&\nodata \\
01541973+2153207&10.18&8.60&9.00&5.29&4.48&4.29&8.57&3.59&10.26&3.61 \\
02255435+3807219&11.90&9.63&\nodata&3.50&2.43&1.92&6.14&1.10&7.54&0.95 \\
03280969+1740466&12.81&11.14&\nodata&7.91&7.02&6.72&11.09&\nodata&\nodata&\nodata \\
03314486+0441437&12.68&11.05&\nodata&6.93&6.02&5.65&9.92&\nodata&\nodata&\nodata \\
03433341+2226132&10.74&9.05&9.69&5.03&4.03&3.80&8.14&3.12&9.72&3.23 \\
03505704+0654325&12.40&10.54&\nodata&6.42&5.47&5.12&9.32&4.32&10.52&\nodata \\
04242179-0231579&10.42&8.78&9.33&5.62&4.78&4.44&8.80&3.96&10.24&3.85 \\
04554504+7959597&10.68&9.04&9.66&5.67&4.87&4.55&8.97&4.04&10.18&3.99 \\
05232068-0434142&11.23&9.04&10.00&3.90&2.80&2.36&\nodata&0.38&6.47&$-0.50$ \\
07145208+6848155&8.85&7.33&\nodata&2.60&1.67&1.39&5.74&0.70&6.92&0.08 \\
07255747+6235276&12.49&10.92&\nodata&7.67&6.83&6.53&10.79&\nodata&\nodata&\nodata \\
07281161+4559261&\nodata&\nodata&7.37&0.65&$-0.38$&$-0.69$&3.43&$-1.59$&4.15&$-0.67$ \\
07491817+2344040&13.37&11.10&8.00&4.12&3.22&2.71&7.16&2.20&8.60&2.05 \\
07504360+4700142&11.83&10.26&\nodata&7.32&6.43&6.17&10.43&\nodata&\nodata&\nodata \\
07525322+3436508&11.34&9.79&\nodata&4.99&3.99&3.71&7.99&2.95&9.38&2.81 \\
07532873+1746498&\nodata&\nodata&8.56&4.82&3.84&3.47&7.76&2.79&9.23&2.83 \\
07565600+3110022&16.96&14.34&\nodata&6.60&5.66&5.13&8.81&3.73&10.11&2.97 \\
08072086+1119187&10.93&9.43&\nodata&7.37&6.52&6.22&10.43&\nodata&\nodata&\nodata \\
08110277+0808214&10.19&8.30&\nodata&5.45&4.58&4.17&8.42&3.52&9.84&3.35 \\
08214282+1717064&\nodata&\nodata&7.50&5.08&4.00&3.59&6.67&1.46&8.07&1.46 \\
09103880+3057472&6.88&4.94&6.08&$-0.71$&$-1.56$&$-1.87$&\nodata&$-3.07$&2.72&$-1.87$ \\
09353643+6909253&10.24&8.78&9.30&5.55&4.69&4.42&8.84&3.84&10.08&3.89 \\
10505517+0429583&12.65&10.90&\nodata&5.36&4.54&4.05&8.28&\nodata&9.46&\nodata \\
11075283+6821591&9.13&7.73&8.10&4.07&3.17&2.77&7.36&2.37&8.81&2.26 \\
12435667+6105357&9.69&7.69&8.87&4.46&3.43&3.02&7.21&2.06&8.39&1.71 \\
13211873+4359136&11.00&9.37&9.78&6.28&5.43&5.16&9.53&\nodata&10.79&\nodata \\
13483402+3159567&13.30&11.80&\nodata&9.03&8.43&8.22&\nodata&\nodata&\nodata&\nodata \\
13520135-0328482&\nodata&\nodata&9.59&5.50&4.76&4.35&8.75&3.78&10.35&\nodata \\
14274800-0304469&10.26&8.77&9.47&5.74&4.93&4.59&9.01&4.05&10.66&\nodata \\
14175111+8349541&10.58&8.81&6.97&3.82&2.90&2.46&6.90&1.83&8.38&1.66 \\
15293218+0011206&\nodata&\nodata&\nodata&7.47&6.70&6.40&10.77&\nodata&\nodata&\nodata \\
15522408-2321075&\nodata&\nodata&\nodata&6.67&5.76&5.38&9.68&4.79&\nodata&\nodata \\
15504662+4828590&8.41&6.50&\nodata&0.74&$-0.14$&$-0.54$&3.09&$-2.12$&3.83&$-2.9$ \\
16213624+5652373&8.50&7.05&7.59&4.71&3.65&3.46&7.66&2.82&9.18&2.64 \\
16370314+0722207&10.50&8.95&9.53&5.50&4.71&4.38&8.64&\nodata&9.98&\nodata \\
17064661-0201039&12.73&10.94&\nodata&7.35&6.36&6.07&\nodata&\nodata&\nodata&\nodata \\
17220711+2349032&10.91&9.37&10.10&5.73&4.98&4.59&8.98&4.09&10.40&3.99 \\
17244818+3355082&12.28&10.73&\nodata&7.73&6.92&6.63&10.98&\nodata&\nodata&\nodata \\
17490067+2142363&12.71&11.22&\nodata&7.78&6.94&6.64&10.95&5.98&\nodata&\nodata \\
18373574+4315237&12.42&10.93&\nodata&7.46&6.61&6.36&10.61&5.86&\nodata&\nodata \\
18513012+4854429&17.02&14.59&10.00&5.82&5.14&4.74&7.99&2.52&8.78&2.01 \\
19364937+5011597&12.21&9.58&8.15&2.25&1.38&0.86&3.94&\nodata&4.87&\nodata \\
19372283+6711199&11.18&9.23&9.70&5.07&4.19&3.76&7.88&2.79&9.30&2.64 \\
19503392+3254509&12.55&9.79&6.80&0.17&$-1.10$&$-1.70$&\nodata&$-4.44$&\nodata&$-1.68$ \\
20313919+0403424&12.43&10.90&\nodata&7.48&6.61&6.25&10.64&\nodata&\nodata&\nodata \\
20532536+0632217&12.78&11.29&\nodata&7.99&7.17&6.90&11.14&\nodata&\nodata&\nodata \\
21102440+0136220&\nodata&\nodata&\nodata&7.60&6.79&6.44&10.71&\nodata&\nodata&\nodata \\
21125578+0932559&\nodata&\nodata&\nodata&7.94&7.09&6.81&11.15&\nodata&\nodata&\nodata \\
22075885+2939562&\nodata&\nodata&\nodata&6.56&5.71&5.39&9.64&4.90&10.87&\nodata \\
22315839+0201206&\nodata&\nodata&\nodata&6.46&5.56&5.21&9.44&\nodata&11.13&\nodata \\
22434918+3355344&\nodata&\nodata&7.82&4.98&4.07&3.79&8.10&3.18&9.60&3.02 \\
22543563+1656308&7.02&5.43&6.47&2.31&1.24&1.04&5.09&$-0.02$&6.63&$-0.08$ \\
23160842+2851492&13.62&12.10&\nodata&9.11&8.32&8.03&\nodata&\nodata&\nodata&\nodata \\
23460974+3446599&12.02&10.53&\nodata&7.49&6.63&6.40&10.70&\nodata&\nodata&\nodata \\
\enddata
\tablecomments{Magnitudes of the S stars in the FAST sample.  We present our derived SDSS $g$ and $r$ magnitudes, as well as the 2MASS magnitudes.  We include IRAS $[12]$ and $[25]$ and AKARI $[S9W]$ and $[L18W]$ magnitudes when available.  Note that the AKARI magnitudes are calculated using the AB system.}
\end{deluxetable}

\begin{deluxetable}{ccccccccc}
\tablewidth{0pc}
\tablecaption{Distances and Absolute Magnitudes\label{tab:plx}}
\tablehead{
\colhead{2MASS Identifier}&\colhead{Parallax}&\colhead{$\sigma{_p}$}&\colhead{Distance}&\colhead{$\sigma{_d}$}&\colhead{abs. g}&\colhead{$\sigma_{g}$}&\colhead{abs. r} &\colhead{$\sigma_{r}$}\\
& (mas) & (mas) & (pc) & (pc) & (mag) & (mag) & (mag) & (mag)}
\startdata
09103880+3057472&8.06&0.98&125&15&1.41&0.33&$-0.53$&0.33 \\
15504662+4828590&3.07&0.75&326&80&0.85&0.56&$-1.06$&0.56 \\
16213624+5652373&2.31&0.62&433&117&0.32&0.61&$-1.13$&0.61 \\
22543563+1656308&3.31&0.93&302&85&0.02&0.64&$-1.97$&0.64 \\
\enddata
\tablecomments{Parallax, distance, and absolute magnitudes for four members of the FAST sample of S Stars
with detectable (3$\sigma$) parallax measurements from Hipparcos.}
\end{deluxetable}

\begin{deluxetable}{clcccccc}
\tablewidth{0pc}
\tablecaption{Proper Motions, and Derived Distance and Magnitude LImits\label{tab:pm}}
\tablehead{
\colhead{2MASS Identifier}&\colhead{Proper
  Motion}&\colhead{$\sigma{_{pm}}$}&\colhead{Distance$^{lim}$}&\colhead{g
  mag}&\colhead{r mag}&\colhead{$M_g^{lim}$}&\colhead{$M_r^{lim}$} \\ 
& (mas/yr) & (mas/yr) & (kpc) & (mag) & (mag) & (mag) & (mag)}
\startdata
00240197+3834373&35.89&2.46&3.22&8.4&6.37&$-$4.14&$-$6.17 \\
01051223+1911522&12.65&0.95&9.13&9.38&7.79&$-$5.42&$-$7.01 \\
03505704+0654325&10.66&2.50&10.8&12.4&10.55&$-$2.77&$-$4.62 \\
04242179-0231579&7.50&1.27&15.4&10.41&8.78&$-$5.53&$-$7.16 \\
07145208+6848155&18.44&1.07&6.26&8.85&7.33&$-$5.13&$-$6.65 \\
09103880+3057472&34.22&0.85&3.38&6.8&4.94&$-$5.84&$-$7.70 \\
09353643+6909253&8.54&1.67&13.5&10.24&8.78&$-$5.42&$-$6.88 \\
11075283+6821591&22.65&1.15&5.10&9.13&7.73&$-$4.41&$-$5.81 \\
12435667+6105357&13.62&1.31&8.48&9.69&7.69&$-$4.95&$-$6.95 \\
13211873+4359136&11.31&2.18&10.2&10.99&9.37&$-$4.06&$-$5.68 \\
14175111+8349541&7.26&1.62&15.9&10.58&8.82&$-$5.43&$-$7.19 \\
14274800-0304469&13.18&1.89&8.77&10.26&8.77&$-$4.45&$-$5.94 \\
15504662+4828590&22.38&1.11&5.16&8.41&6.5&$-$5.15&$-$7.06 \\
16213624+5652373&12.32&1.14&9.37&8.5&7.05&$-$6.36&$-$7.81 \\
19372283+6711199&9.75&2.35&11.8&11.18&9.24&$-$4.19&$-$6.13 \\
22543563+1656308&15.75&0.99&7.33&7.02&5.43&$-$7.31&$-$8.90 \\
\enddata
\tablecomments{Stars from the FAST sample with significant proper
  motions and derived quantities.  Distances are upper limits,
  as described in \S\,\ref{sec:motion2}, derived by assuming
  $v_{trans}<545$km/s.  Absolute magnitudes are the corresponding 
  lower (brightest) limits.  We can immediately see that these stars
  could all be giants.  We do not present errors in  the distance and
  magnitude calculations, since these are rough upper 
  limits, rather than rigorous calculations of the actual expected
  magnitude.} 
\end{deluxetable}

\vfill
\eject

\begin{thebibliography}{}
\bibitem[Ake(1979)]{1979ApJ...234..538A} Ake, T.~B.\,1979, \apj, 234, 538 
\bibitem[Bergeat et al.(2002)]{Bergeat02} Bergeat, J., Knapik, A., \& Rutily, B.\,2002, \aap, 385, 94 
\bibitem[Bodenheimer(1965)]{1965ApJ...142..451B} Bodenheimer, P.\,1965, \apj, 142, 451 
\bibitem[Carpenter(2001)]{2001AJ....121.2851C} Carpenter, J.~M.\,2001, \aj, 121, 2851 
\bibitem[Covey et al.(2007)]{2007AJ....134.2398C} Covey, K.~R., et al.\,2007, \aj, 134, 2398 
\bibitem[Dahn et al.(1977)]{1977ApJ...216..757D} Dahn, C.~C., Liebert, J., Kron, R.~G., Spinrad, H., \& Hintzen, P.~M.\,1977, \apj, 216, 757 
\bibitem[Downes et al.(2004)]{2004AJ....127.2838D} Downes, R.~A., et al.\,2004, \aj, 127, 2838 
\bibitem[Fukugita et al.(1996)]{1996AJ....111.1748F} Fukugita, M., Ichikawa, T., Gunn, J.~E., Doi, M., Shimasaku, K., \& Schneider, D.~P.\,1996, \aj, 111, 1748 
\bibitem[Gray and Corvally (2009)]{Gray09} Gray, R.~O., Corbally,  C.~J.\,Stellar Spectral Classification, \,Princeton University Press, \,2009
\bibitem[Green et al.(1991)]{Green91} Green, P.~J., Margon, B., 
\& MacConnell, D.~J.\,1991, \apjl, 380, L31 
\bibitem[Green et al.(1994)]{1994ApJ...434..319G} Green, P.~J., Margon, B., Anderson, S.~F., \& Cook, K.~H.\,1994, \apj, 434, 319 
\bibitem[Green et al.(1992)]{1992ApJ...400..659G} Green, P.~J., Margon, B., Anderson, S.~F., \& MacConnell, D.~J.\,1992, \apj, 400, 659 
\bibitem[Groenewegen(1993)]{1993A&A...271..180G} Groenewegen, M.~A.~T.\,1993, \aap, 271, 180 
\bibitem[Herwig(2005)]{2005ARA&A..43..435H} Herwig, F.\ 2005, \araa, 43, 435 
\bibitem[H{\o}g et al.(1997)]{1997A&A...323L..57H} H{\o}g, E., et al.\,1997, \aap, 323, L57 
\bibitem[Houdashelt et al.(2000)]{2000AJ....119.1424H} Houdashelt, M.~L., Bell, R.~A., Sweigart, A.~V., \& Wing, R.~F.\,2000, \aj, 119, 1424 
\bibitem[Iben \& Renzini(1983)]{1983ARA&A..21..271I} Iben, I., Jr., \& Renzini, A.\ 1983, \araa, 21, 271 
\bibitem[Ishihara et al.(2010)]{2010A&A...514A...1I} Ishihara, D., et al.\,2010, \aap, 514, A1 
\bibitem[Jorissen et al.(1993)]{1993A&A...271..463J} Jorissen, A., Frayer, D.~T., Johnson, H.~R., Mayor, M., \& Smith, V.~V.\,1993, \aap, 271, 463 
\bibitem[Keenan \& Boeshaar(1980)]{1980ApJS...43..379K} Keenan, P.~C., \& Boeshaar, P.~C.\,1980, \apjs, 43, 379 
\bibitem[Keenan(1954)]{1954ApJ...120..484K} Keenan, P.~C.\,1954, \apj, 120, 484 
\bibitem[Margon, B., et al.(2002)]{Margon02} Margon, B. et al., 2002 AJ, 124, 1651
\bibitem[Murakami et al.(2007)]{2007PASJ...59S.369M} Murakami, H., et  al.\,2007, \pasj, 59, 369 
\bibitem[Perryman et al.(1997)]{1997A&A...323L..49P} Perryman, M.~A.~C., et al.\,1997, \aap, 323, L49 
\bibitem[Plez et al.(2003)]{2003IAUS..210P..A2P} Plez, B., van Eck, S., Jorissen, A., Edvardsson, B., Eriksson, K., \& Gustafsson, B.\,2003, Modelling of Stellar Atmospheres, 210, 2P 
\bibitem[Sirko et al.(2004)]{2004AJ....127..914S} Sirko, E., et al.\,2004, \aj, 127, 914 
\bibitem[Skrutskie et al.(2006)]{2006AJ....131.1163S} Skrutskie, M.~F., et al.\,2006, \aj, 131, 1163 
\bibitem[Smith et al.(2002)]{2002AJ....123.2121S} Smith, J.~A., et al.\,2002, \aj, 123, 2121 
\bibitem[Smith et al.(2007)]{2007MNRAS.379..755S} Smith, M.~C., et al.\,2007, \mnras, 379, 755 
\bibitem[Stephenson(1984)]{1984PW&SO...3....1S} Stephenson, C.~B.\,1984, Publications of the Warner \& Swasey Observatory, 3, 1 
\bibitem[Valdes F., et al.(2004)]{Valdes04} Valdes F., Gupta R., Rose  J.A., Singh H.P., Bell D.J. 2004, Astrophys. J. Suppl. Ser., 152, 251-259 
\bibitem[Van Eck \& Jorissen(2000)]{2000A&A...360..196V} Van Eck, S., \& Jorissen, A.\,2000, \aap, 360, 196 
\bibitem[Van Eck et al.(2010)]{2010arXiv1011.2092V} Van Eck, S., et al.\ 
2010, arXiv:1011.2092 
\bibitem[Vyssotsky(1947)]{1947PMcCO...9..197V} Vyssotsky, A.~N.\,1947, Publications of the Leander McCormick Observatory, 9, 197 
\bibitem[Wyckoff \& Clegg(1978)]{1978MNRAS.184..127W} Wyckoff, S., \& Clegg, R.~E.~S.\,1978, \mnras, 184, 127 
\bibitem[Yang et al.(2006)]{2006AJ....132.1468Y} Yang, X., Chen, P., Wang, J., \& He, J.\,2006, \aj, 132, 1468 
\end{thebibliography}
\end{document}